\documentclass[twoside]{article}
\usepackage{multicol,amsmath}  

\evensidemargin=\oddsidemargin \oddsidemargin=0.5cm \evensidemargin=0.5cm  
\font\twbf=cmbx10 at 11pt   
\font\bsl=cmbxsl10 at 11pt 
\font\bfit=cmbxti10   

\catcode`\@=11
\newcounter{sectc}\newcounter{subsectc}\newcounter{subsubsectc}
\newcommand{\sect}[1] {\vspace{0.3cm}\addtocounter{sectc}{1} 
\setcounter{subsectc}{0}\setcounter{subsubsectc}{0}\noindent 
      {\twbf\thesectc\ \  #1}\par\vspace{0.2cm}}
\newcommand{\subsect}[1] {\vspace{0.3cm}\addtocounter{subsectc}{1} 
          \setcounter{subsubsectc}{0}\noindent 
          {\bf\thesectc.\thesubsectc\ \ \bfit  #1}\par\vspace{0.15cm}}

\newcommand{\nonumsect}[1] {\vspace{0.3cm}\noindent{\twbf #1}
          \par\vspace{0.2cm}}
\newcounter{appendixc}
\newcounter{subappendixc}[appendixc]
\newcounter{subsubappendixc}[subappendixc]

\renewcommand{\appendix}[1] {\vspace{0.3cm}
        \refstepcounter{appendixc}
        \setcounter{figure}{0}
        \setcounter{table}{0}
        \setcounter{equation}{0}
        \renewcommand{\thefigure}{\Alph{appendixc}.\arabic{figure}}
        \renewcommand{\thetable}{\Alph{appendixc}.\arabic{table}}
        \renewcommand{\theappendixc}{\Alph{appendixc}}
        \renewcommand{\theequation}{\Alph{appendixc}.\arabic{equation}}
        \noindent{\bf Appendix \theappendixc\ \ #1}\par\vspace{0.2cm}}

\if@twoside \def\ps@headings{\let\@mkboth\markboth
\def\@oddfoot{}\def\@evenfoot{}\def\@evenhead{\rm\footnotesize \thepage\hfil 
\footnotesize
\leftmark}\def\@oddhead{\hbox{}\footnotesize \rightmark \hfil
\rm\thepage}\def\sectionmark##1{\markboth {\uppercase{\ifnum \c@secnumdepth
>\z@
 \thesection\hskip 1em\relax \fi ##1}}{}}\def\subsectionmark##1{\markright
{\ifnum \c@secnumdepth >\@ne
 \thesubsection\hskip 1em\relax \fi ##1}}}
\else \def\ps@headings{\let\@mkboth\markboth
\def\@oddfoot{}\def\@evenfoot{}\def\@oddhead{\hbox {}\footnotesize \rightmark 
\hfil\rm\footnotesize\thepage}\def\sectionmark##1{\markright 
{\uppercase{\ifnum \c@secnumdepth
>\z@
 \thesection\hskip 1em\relax \fi ##1}}}}
\fi
\def\ps@myheadings{\let\@mkboth\@gobbletwo
\def\@oddhead{\hbox{}\footnotesize\rightmark \hfill\rm\footnotesize\thepage}
\def\@oddfoot{}
\def\@evenhead{\rm\footnotesize\thepage 
 \hfill \footnotesize\leftmark\hbox{}}\def\@evenfoot{}
\def\sectionmark##1{}\def\subsectionmark##1{}}
\def\d{\hspace*{.8\p@}\mathstrut\text{d}\hspace*{.6\p@}} 
\def\e{\mathstrut\,\text{e\/}\hspace*{.6\p@}} 
\def\i{\hspace*{.6\p@}\mathstrut\text{i}\hspace*{.6\p@}} 
\catcode`\@=13
\def\Td#1#2#3#4#5{{\thispagestyle{empty}
\protect\headheight0pt\protect\headsep0pt\protect\vspace*{-2.2cm}
{\flushleft\parbox{181mm}{\footnotesize Commun.\ Theor.\ Phys.~(Beijing, 
China) {\bf #1}{~(#2)~}{pp~#3}\\[-0.7mm]
\copyright\hspace*{3.5pt} International Academic Publishers\hfill Vol.~{#4},
No.~{#5}}\\[-1.4mm]
\begin{table}[h]\hfill\null\hfill\hrule\vskip.4mm\hrule\end{table} }}}
\def\no#1{\rlap{\protect\rule[-0.25 true cm]{\textwidth}{0.03 true cm}}%
No.~{#1}\hfill}
\def\vo#1{\hfill{\ignorespaces Vol.~{#1}%
\llap{\protect\rule[-0.25 true cm]{\textwidth}{0.03 true cm}}}}

\def\rd{\protect\footnotesize}
\def\pacs#1{\protect\vglue6pt%
\noindent\begin{minipage}{170mm}{\bf PACS numbers: }\rm #1\end{minipage}}
\def\key#1{\leftskip=0.5cm\vglue0pt\noindent\hspace*{-1pt}\protect
\begin{minipage}{170mm}\begin{minipage}[t]{21mm}{\bf Key words:}\end{minipage}
\hfill\begin{minipage}[t]{148mm}\rm\baselineskip=12pt #1\end{minipage}
\end{minipage}\par\leftskip=0cm}
\def\title#1{\begin{flushleft} 
\large\bf\protect\baselineskip=17pt #1\end{flushleft}}
\def\author#1{\leftskip=0.5cm\noindent\begin{minipage}{170mm}
\normalsize\hspace*{-4.5pt}#1\end{minipage}\par\vglue4pt} 
\def\address#1#2{\leftskip=0.5cm\noindent\begin{minipage}{170mm}\parindent=-4.5pt   
${}^{#1}$\protect\small\baselineskip=10pt #2\end{minipage}\par\vglue2pt} 
 
\def\date#1{\leftskip=0.5cm\vglue4pt\noindent\begin{minipage}{170mm}
\normalsize\hspace*{-4.5pt}#1\end{minipage}\par\vglue4pt} 
\def\abstract#1#2{\leftskip=0.5cm\vglue4pt\noindent\hspace*{-4.5pt}
\begin{minipage}{170mm}\small\sl\baselineskip=11pt{\bsl Abstract\ } #1\par
\pacs{\normalsize#2}\end{minipage}\par\vglue2pt\leftskip=0cm}

\def\ruledown{\hfill\noindent{\lower.38cm\hbox{\rule{0.6pt}{0.4cm}}\rule{9.05cm}
              {0.6pt}}\vspace*{-0.6cm}}
\def\refrule{\vglue5pt\centerline{\hbox to 10cm{\hrulefill}}\vglue10pt}
\def\wen#1{$^{[#1]}$}
\def\noa#1{\noalign{\vskip#1pt}}
\def\alw{\allowdisplaybreaks}
\def\ssc#1{\scriptscriptstyle{#1}}

\def\mylistlabel#1{#1}
\newenvironment{mylabellist}[1]
            {\hbadness=15000
\begin{list}{(\mylistlabel)}
    {\itemindent0pt 
    \leftmargin2\parindent 
    \setlength{\parsep}{0pt}
             \setlength{\itemsep}{0pt}
             
              \settowidth{\labelwidth}{#1}
             }}{\end{list}}

\begin{document}
\hoffset=-1.4cm  \voffset=0.4cm
\abovedisplayskip=8pt plus 1pt minus 1pt
\belowdisplayskip=8pt plus 1pt minus 1pt
\parskip=0.2pt plus.5pt minus0.2pt

\setcounter{footnote}{0}
\setcounter{page}{11} 
\Td{36}{2001}{11--18}{36}{1, July 15, 2001} 

\title{A Note on Symplectic Algorithm}
\author{GUO Han-Ying, LI Yu-Qi and WU Ke}
\address{}{Institute of Theoretical Physics, Academia Sinica,
P.O.\ Box 2735, Beijing 100080, China}

\date{(Received March 23, 2001)}
\abstract{We present the symplectic algorithm in the 
Lagrangian formalism for the Hamiltonian systems by virtue of the
noncommutative differential calculus  with respect to the discrete time 
and the Euler--Lagrange cohomological concepts.  We also
show that the trapezoidal integrator is symplectic in certain
sense.}{02.60.Li, 11.10.Ef}
\key{symplectic algorithm, Lagrangian formalism, Euler-Lagrange cohomology}

\vspace*{0.4cm}

\baselineskip=13.6pt plus.5pt minus.2pt

\begin{multicols}{2}
\sect{Introduction}

It is well known that the symplectic  algorithm\wen{1,2} for the
finite dimensional Hamiltonian systems are  very powerful and successful in 
numerical calculations  in comparison with  other various non-symplectic 
computational  schemes  since the symplectic schemes preserve the symplectic 
structure in certain sense.  On the other hand, the Lagrangian formalism is 
quite useful for the Hamiltonian systems. Since  both are important at least 
in the equal footing, it should not be useless to establish the symplectic 
algorithms in the Lagrangian formalism. As a matter of fact, the Lagrangian 
formalism is more or less earlier to be generalized to the infinite 
dimensional systems such as classical field theory.

In this note we present the symplectic geometry and symplectic algorithm in 
the Lagrangian formalism in addition to the Hamiltonian one for the finite 
dimensional Hamiltonian systems with the help of the Euler--Lagrange (EL)
cohomological concepts introduced very recently by the authors in Ref.~[5]. 

In the  course of numerical calculation, the ``time''  $t \in R$ is always 
discretized, usually with equal spacing $h=\Delta t$,
$$
t\in R  \rightarrow t\in  {\it T}=\{ (t_k , t_{k+1}=t_k+h,  ~~k \in Z)\}\,.
$$
It is well known that the differences of functions on $T$ with
respect to $T$ do not
obey the Leibniz law. In order to show that the symplectic structure at
different moments $t_k$ is preserved, some well-established differential
calculus should be employed.
This implies that some noncommutative differential calculus (NCDC) 
on $T$ and the function space on it should be used
 even for the well-established symplectic algorithms. In this note we employ
this simple NCDC.\wen{3,4} We also show that the trapezoidal integrator 
is symplectic in certain sense. Finally, we end with some remarks.

\sect{The Necessary and Sufficient Condition 
for\newline\phantom{2\ \ }Symplectic Preserving Law}
\vspace*{-0.15cm}

In this section, we first recall some well-known contents in the
Lagrangian formalism for the finite dimensional Hamiltonian systems. 
We employ the ordinary calculus to show that the symplectic structure 
is preserved by introducing the EL cohomological concepts such as
the EL one-forms, the null EL one-form, the coboundary EL one-form and 
the EL condition and so forth.\wen{5}
It is important to emphasize that the symplectic structure-preserving is
in the function space in general rather than in the solution space of the
EL equation only. The reason will be explained later.

Let time $t\in R$ be the base manifold, $M=M^{n}$ the 
configuration space on $t$,
$$q=[q^1(t), \cdots, q^n(t)]^T$$ 
the (canonical) coordinates on it, $T$ the
transport, $TM$ the tangent bundle of $M$ with coordinates 
$$
(q, \dot q)=([q^1(t), \cdots, q^n(t)]^T,[{\dot q}^1(t), \cdots, 
{\dot q}^n(t)]^T)\,,$$
$F(TM)$ the function space on $TM$.

The Lagrangian of the systems under consideration is $L(q^i, {\dot q^j})$
with the well-known EL equation  from the variational principle
\begin{align}
\frac {\partial L}{\partial {q^i}}-\frac{\d}{\d t}
\frac{\partial L}{\partial {\dot q^i}}=0\,.
\end{align}
Let us introduce the EL one-form\wen{5}
\begin{align}
E(q^i, {\dot q^j}):\;=\Biggl\{\frac{\partial L}{\partial
{q^i}}-\frac{\d}{\d t}\frac{\partial L}{\partial {\dot q^i}}\Biggr\}
\d q^i\,.
\end{align}
It is clear that the EL equation is given by the null EL one-form,
$$
E(q^i, {\dot q^j})=0\,,
$$
which is a special case of the coboundary EL one-forms
\begin{align}
E(q^i, {\dot q^j})=\d\alpha (q^i, {\dot q^j})\,,
\end{align}
where $\alpha (q^i, {\dot q^j})$ is an arbitrary function of 
$ (q^i, {\dot q^j})$  in the function space $F(TM)$.

Taking the exterior derivative d of the Lagrangian, we get
\begin{align}
\d L(q^i, {\dot q^j})=E(q^i, {\dot q^j})
+\frac{\d}{\d t} \theta\,,
\end{align}
where $\theta$ is the canonical one-form defined by
\begin{align}
\theta=\frac{\partial L} {\partial {\dot q^i}} \d q^i\,.
\end{align}
Making use of nilpotency of d, 
$$
\d^2L(q, {\dot q})=0\,,
$$  
it follows that iff the EL one-form is closed 
with respect to d, i.e.,
\begin{align}
\d E(q^i, {\dot q^j})=0\,,
\end{align}
which is called the EL condition,\wen{5}  the symplectic conservation 
law with respect to $t$ holds,
\begin{align}
\frac{\d}{\d t} \omega = 0\,,
\end{align}
where the symplectic structure $\omega$ is given by
\begin{align}
\omega=\d\theta =\frac{\partial^2 L}{\partial{\dot q^i}
{\partial q^j}}\d q^i \wedge \d q^j
+\frac {\partial^2 L}{\partial {\dot q^i}{\partial {\dot q^j}}} 
\d q^i \wedge \d{\dot q}^j\,.
\end{align}

It is important to note that although the null EL one-form and the coboundary
EL one-forms  satisfy the EL condition and they   cohomologically 
trivial, this does not mean that the closed EL one-forms are always exact. 
As a matter of
fact, the equation (8) shows that the EL one-form is not always exact
since the canonical one-form is not trivial in general. 
In addition, it is also important to note that  
the $q^i(t)$, $i=1, \ldots, n$ in the EL condition are still in the function
space in general rather than in the solution space of the equation only. 
This means that the symplectic two-form $\omega$  is conserved with respect 
to $t$ with the closed EL condition in general
rather than in the solution space only.

In order to transfer to the Hamiltonian formalism, we introduce the canonical
momentum 
\begin{align}
p_j=\frac {\partial L} {\partial \dot{q}^i}\,,
\end{align}
and take a Legendre transformation to get the Hamiltonian function
\begin{align}
H(q^i,p_j)=p_k {\dot q}^k -L(q^i, {\dot q}^j)\,.
\end{align}
Then the EL equation becomes the canonical equations as follows:
\begin{align}
 \dot q^i=\frac {\partial H} {\partial {p_i}}\,, \qquad 
 \dot p^j=-\frac {\partial H} {\partial { q^j}}\,.
\end{align} 
It is clear that a pair of the EL one-forms should be introduced now, 
\begin{align}
&E_1(q^i, p_j)=\Biggl( \dot{q}^j-\frac{\partial H}{\partial p_j}\Biggr)\d p_j,
\nonumber\\\noa2
&E_2(q^i, p_j)=-\Biggl(\dot{p}_j+\frac{\partial H}{\partial q^j}\Biggr)
\d q^j\,.
\end{align}
In terms of $z^T={(q^i, \ldots, q^n, p_1, \ldots, p_n)}$, the canonical 
equations and the EL one-form become
\begin{align}
&\dot z=J^{-1}\nabla_z H\,,
\\noa2
&E(z)=\d z^T(Jz-\nabla_z H)\,.
\end{align}
Now it is straightforward to show that the symplectic structure-preserving 
law
\begin{align}
& \frac{\d}{\d t}\omega=0\,, \nonumber\\\noa2 
& \omega=\d z^T \wedge J\d z
\end{align} 
holds {\bfit if and only if} the (closed) EL condition is satisfied
\begin{align}
\d E(z)=0\,.
\end{align}

\sect{The Necessary and Sufficient Condition 
for\newline\hphantom{3\ \ }Discrete Symplectic Preserving Law}
\vspace*{-0.15cm}

Now we consider the symplectic structure-preserving of symplectic algorithm
in the Lagrangian formalism.
As was mentioned above, in the  course of numerical calculation,
 the ``time''  $t \in R$ is discretized with equal spacing $h=\Delta t $,
\begin{align}
t\in R  \rightarrow t\in   T=\{ (t_k , t_{k+1}=t_k+h,  ~~k \in
Z)\}\,.
\end{align}
At the   moment  $t_k$, the coordinates of the space 
$$
M_k^n \in M_T^{n}=\{\cdots M_1^n \times \cdots \times M_k^n \cdots \} 
$$ 
are denoted by $q^{(k)}$, the  symplectic structure by
\begin{align}
\omega^{(k)}={\d q_t^{(k)}}^T \wedge \d q^{(k)}\,,
\end{align}
and  $q_t^{(k)}$ now is the (forward-)difference of $q^{(k)}$,
\begin{align}
\Delta_tq^{(n)}:\;=\partial_tq^{(n)}=q_t^{(k)}=\frac 1 h ( q^{(k+1)}- q^{(k)})\,.
\end{align}
Now the EL equation becomes the  difference discrete Euler--Lagrange (DEL)
equation which can be derived from the difference discrete variational
principle,\wen{5}
\begin{align}
\frac {\partial L_D^{(k)}}{\partial q^{i(k)}} 
-\Delta_t\Biggl(\frac {\partial L_D^{(k-1)}}{\partial 
{q^i_t}^{(k-1)}}\Biggr)=0\,.
\end{align}
\end{multicols}

\abovedisplayskip=6.5pt plus 1pt minus 1pt
\belowdisplayskip=6.5pt plus 1pt minus 1pt

Now we consider the difference discrete symplectic structure and its preserving
property. Taking the exterior derivative d on  $T^*(M_{T_D}^n)$, we get
$$
\d L_D^{(k)}= \frac{\partial L_D^{(k)}}{\partial{q^i}^{(k)}}\d{q^i}^{(k)}
+\frac {\partial L_D^{(k)}} {\partial {q^i_t}^{(k)}}\d {q^i_t}^{(k)}\,.
$$
By means of the modified Leibniz law with respect to  $\Delta_t$ and 
introducing the DEL one-form,
\begin{align}
E_D^{(k)}({q^i}^{(k)}, {q^j_t}^{(k)}):\;=\Biggl\{
\frac{\partial L_D^{(k)}}{\partial {q^i}^{(k)}} 
-\Delta_t\Biggl(\frac{\partial L_D^{(k-1)}}{\partial
{q^i_t}^{(k-1)}}\Biggr)\Biggr\}
\d{q^i}^{(k)},
\end{align}
we have
\begin{align}
\d L_D^{(k)}=E_D^{(k)}+\Delta_t \theta_{\ssc D}^{(k)}\,,
\end{align}
where $\theta_{\ssc D}^{(k)}$ is the discrete canonical one-form,
\begin{align}
\theta_{\ssc D}^{(k)}=\frac{\partial L_D^{(k-1)}}{\partial{q^i_t}^{(k-1)}}
\d{q^i}^{(k)}\,,
\end{align}
then there exists the following discrete symplectic 
two-form on $T^*(M_{T_D}^n)$,
\begin{align}
\omega_{\ssc D}^{(k)}=\d\theta_{\ssc D}^{(k)}
=\frac {\partial^2 L_D^{(k-1)}}{\partial {q^i_t}^{(k-1)} 
\partial {q^j}^{(k-1)}} \d{q^j}^{(k-1)} \wedge \d{q^i}^{(k)}
+\frac{\partial^2 L_D^{(k-1)}}{\partial 
{q^i_t}^{(k-1)} \partial {q^j_t}^{(k-1)}}\d{q^j_t}{(k-1)} \wedge 
\d{q^i}^{(k)}\,.
\end{align}
It is easy to see that the null DEL one-form $E_D^{(k)}=0$ 
gives rise to the DEL equation and it is a special case of the coboundary
DEL one-forms
\begin{align}
E_D^{(k)}=\d\alpha_{\ssc D}^{(k)}({q^i}^{(k)}, {q^j_t}^{ (k)})\,,
\end{align}
where $\alpha_{\ssc D}^{(k)}({q^i}^{(k)}, {q^j_t}^{ (k)})$ is an arbitrary 
function of $({q^i}^{(k)}, {q^j_t}^{ (k)})$.

Finally,  due to the nilpotency of d on  $T^*(M_{T_D}^n)$ 
it is easy to prove  from  Eq.~(22) that iff the DEL one-form satisfies 
what is called the DEL condition
\begin{align}
\d E_D^{(k)}=0\,, 
\end{align}
i.e., the DEL one-form is closed,  the discrete 
(difference) symplectic structure-preserving law holds,
\begin{align}
\Delta_t \omega_{\ssc D}^{(k)}=0\,.
\end{align}

Similar to the continuous case, the closed DEL one-forms are not always 
exact and this difference discrete  symplectic structure-preserving law 
is held in function space in general rather than  in  solution 
space only.

Let us consider the following DEL equation
\begin{align}
q^{(k)}-2q^{(k+1)}+q^{(k+2)}=-h^2\frac{\partial L}{\partial q}(q^{(k+1)})\,.
\end{align}
Introducing the DEL one-form\wen{5} 
\begin{align}
E_D^{(k+1)}:\;=\d(q^{T(k+1)})\Bigl\{q^{(k)}-2q^{(k+1)}
+q^{(k+2)}-h^2 \frac{\partial L}{\partial q} (q^{(k+1)})\Bigr\}\,,
\end{align}
the null DEL one-form is corresponding to the DEL equation and the DEL 
condition directly gives rise to
$$ 
\d q^{T(k+2)} \wedge \d q^{(k+1)}= \d q^{T(k+1)}\wedge \d q^{(k)}\,.
$$
It follows that
\begin{align}
\Delta_t \omega^{(k)}
 =\frac1h [\d q_t^{T(k+1)} \wedge \d q^{(k+1)}- \d q_t^{T(k)} 
\wedge \d q^{(k)}]=0\,. 
\end{align}
This means that the (forward-)difference scheme is symplectic.
It can be proved that the scheme with respect to the (backward-)difference 
of $q^{(n)}$,
\begin{align}
\Delta_tq^{(k)}=q_t^{(k)}=\frac 1 h ( q^{(k)}- q^{(k-1)})
\end{align} 
is also symplectic as well.

\sect{On the Symplectic Schemes} 
\vspace*{-0.15cm}

We now show some well-known symplectic schemes in the Lagrangian formalism
for the Hamiltonian systems.

\subsect{The Euler Mid-point Scheme for Separable Hamiltonian Systems}

The well-known Euler mid-point scheme for the separable Hamiltonian systems
is as follows:
\begin{align}
q^{(n+1)}-q^{(n)}=\frac{h}{2}(p^{(n+1)}+p^{(n)})\,,
\qquad 
p^{(n+1)}-p^{(n)}=-hV_q\Bigl(\frac{q^{(n+1)}+q^{(n)}}{2}\Bigr)\,,
\end{align}
from which it follows that
\begin{align}
&(p^{(n+1)}-p^{(n)})+(p^{(n)}-p^{(n-1)})
=-hV_q\Bigl(\frac{q^{(n+1)}+q^{(n)}}{2}\Bigr)
-hV_q\Bigl(\frac{q^{(n)}+q^{(n-1)}}{2}\Bigr)\,,
\\\noa2
&\frac{2}{h}(q^{(n+1)}-q^{(n)}-q^{(n)}+q^{(n-1)})
=-hV_q\Bigl(\frac{q^{(n+1)}+q^{(n)}}{2}\Bigr)
-hV_q\Bigl(\frac{q^{(n)}+q^{(n-1)}}{2}\Bigr)\,.
\end{align}
Now it is easy to get the Euler mid-point scheme in the Lagrangian formalism,
\begin{align}
q^{(n+1)}-2q^{(n)}+q^{(n-1)}=-\frac {h^2}{2}\Bigl[V_q\Bigl(
\frac{q^{(n+1)}+q^{(n)}}{2}\Bigr)+V_q\Bigl(\frac{q^{(n)}
+q^{(n-1)}}{2}\Bigr)\Bigr]\,.
\end{align}
In order to show that it is symplectic, we first introduce the DEL one-form as 
follows: 
\begin{align}
E_D^{(n)}=\d q^{T(n)}\Bigl\{q^{(n+1)}-2q^{(n)}+q^{(n-1)}
+\frac {h^2}{2}\Bigl[V_q\Bigl(\frac{q^{(n+1)}+q^{(n)}}{2}\Bigr)
+V_q\Bigl(\frac{q^{(n)}+q^{(n-1)}}{2}\Bigr)\Bigr]\Bigr\}\,.
\end{align}
Then the DEL condition gives rise to
\begin{align}
\d q^{(n+1)}\wedge \d q^{T(n)}+\d q^{(n-1)}\wedge \d q^{T(n)}
=-\frac{h^2}{4}V_{qq}^{(n+1/2)}\d q^{(n+1)}\wedge \d q^{T(n)}
-\frac{h^2}{4}V_{qq}^{(n-1/2)}\d q^{(n-1)}\wedge \d q^{T(n)}\,,
\end{align}
where 
$$ 
V^{(n+1/2)}=V\Bigl(\frac{q^{(n+1)}+q^{(n)}}{2}\Bigr).
$$
That is
\begin{align}
\Bigl(1+\frac{h^2}{4}V_{qq}^{(n+1/2)}\Bigr)\d q^{T(n)}\wedge \d q^{(n+1)}=
\Bigl(1+\frac{h^2}{4}V_{qq}^{(n-1/2)}\Bigr)\d q^{T(n)}\wedge \d q^{(n-1)}\,.
\end{align}
Now it is easy to prove that
\begin{align}
\d p^{T(n+1)}\wedge \d q^{(n+1)}
=\Bigl(1+\frac{h^2}{4}V_{qq}^{(n+1/2)}\Bigr)\d q^{T(n+1)}
\wedge \d q^{(n)}\,.
\end{align}
Therefore, the Euler mid-point scheme is symplectic.

\subsect{The Euler Mid-point Scheme for Generic Hamiltonian Systems}

For the general Hamiltonian $H$, the similar preserved symplectic form 
can also be given. Let us start with
\begin{align}
&q^{(n+1)}-q^{(n)}=hH_p\Bigl(\frac {p^{(n+1)}+p^{(n)}}{2},
\frac {q^{(n+1)}+q^{(n)}}{2}\Bigr)
\nonumber\\\noa2
&p^{(n+1)}-p^{(n)}=-hH_q\Bigl(\frac{p^{(n+1)}+p^{(n)}}{2},
\frac {q^{(n+1)}+q^{(n)}}{2}\Bigr)\,.
\end{align}

Introduce a pair of  DEL one-forms
\begin{align}
&E_D^{(n)}(q)=\d q^{T(n)}\Bigl\{q^{(n+1)}-q^{(n)}-hH_p
\Bigl(\frac {p^{(n+1)}+p^{(n)}}{2},
\frac {q^{(n+1)}+q^{(n)}}{2}\Bigr)\Bigr\}\,,
\nonumber\\\noa2
&E_D^{(n)}(p)=\d p^{T(n)}\Bigl\{p^{(n+1)}-p^{(n)}+hH_q\Bigl(
\frac {p^{(n+1)}+p^{(n)}}{2},
\frac {q^{(n+1)}+q^{(n)}}{2}\Bigr)\Bigr\}\,.
\end{align}

The DEL conditions for the pair of the DEL one-forms now read
\begin{align}
\d( E_D^{(n)}(q)+ E_D^{(n)}(p))=0\,.
\end{align}
From these conditions it follows that 
\begin{align}\alw
&(\d q^{(n)})^T\wedge\Bigl[\frac{1}{2}hH_{qq}^{(n+1/2)}
+2\Bigl(1+\frac{1}{2}hH_{qp}^{(n+1/2)}\Bigr)(hH_{pp}^{(n+1/2)})^{-1}
\Bigl(1-\frac{1}{2}hH_{pq}^{(n+1/2)}\Bigr)\Bigr]\d q^{(n+1)}
\nonumber\\\noa2
&\qquad=(\d q^{(n-1)})^T\wedge\Bigl[\frac{1}{2}hH_{qq}^{(n-1/2)}
+2\Bigl(1+\frac{1}{2}hH_{qp}^{(n-1/2)}\Bigr)(hH_{pp}^{(n-1/2)})^{-1}
\Bigl(1-\frac{1}{2}hH_{pq}^{(n-1/2)}\Bigr)\Bigr]\d q^{(n)}\,.
\end{align}
This shows that the following two-form in $(\d q^{(k)})$ is preserved,  
\begin{align}
(\d q^{(n-1)})^T\wedge\Bigl[\frac{1}{2}hH_{qq}^{(n-1/2)}
+2\Bigl(1+\frac{1}{2}hH_{qp}^{(n-1/2)}\Bigr)(hH_{pp}^{(n-1/2)})^{-1}
\Bigl(1-\frac{1}{2}hH_{pq}^{(n-1/2)}\Bigr)\Bigr]\d q^{(n)}\,.
\end{align}

It can be shown that it is nothing but the preserved symplectic structure,
\begin{align}
&2(\d p^{(n+1)})^T\wedge \d q^{(n+1)}
=-(\d q^{(n)})^T\wedge \Bigl[\frac{1}{2}hH_{qq}^{(n+1/2)}
\nonumber\\\noa2
&\phantom{2(\d p^{(n+1)})^T\wedge \d q^{(n+1)}=}
+2\Bigl(1+\frac{1}{2}hH_{qp}^{(n+1/2)}\Bigr)(hH_{pp}^{(n+1/2)})^{-1}
\Bigl(1-\frac{1}{2}hH_{pq}^{(n+1/2)}\Bigr)\Bigr] \d q^{(n+1)}\,.
\end{align}
In terms of $z^{\ssc T}=(q^{\ssc T}, p^{\ssc T})$, the mid-point 
scheme can be expressed as
\begin{align}
\Delta_t z^{(k)}=J^{-1}\nabla_z {H_D}^{(k)}\Bigl(\frac12 (z^{(k+1)}+z^{(k)})
\Bigr)\,.
\end{align}
The DEL one-form for the scheme at $t_k$ now becomes
\begin{align}
{E_{D1}}^{(k)}=\frac12 \d({z^{(k+1)}+z^{(k)})}^T\Bigl\{ J \Delta_t z^{(k)}
-\nabla_z {H_D}^{(k)}\Bigl(\frac12 (z^{(k+1)}+z^{(k)})\Bigr)\Bigr\}\,.
\end{align}
It is now straightforward to show that the symplectic structure-preserving 
law
$$
\Delta_t (\d{z^{(k)}}^T\wedge J \d z^{(k)})=0
$$
holds {\bfit if and only if} the DEL form is closed.

\subsect{The High Order Symplectic Schemes}

Similarly, it can be  checked that the high order symplectic schemes
preserve also some two-forms in $\d q^{(k)}$ which are in fact the
symplectic structures. Let us consider two examples for this point.

The first one is proposed by Feng {\it et al.} 
in terms of generating function.\wen{6} The scheme is as follows:
\begin{align}
z^{(n+1)} = z^{(n)}+hJ^{-1}\nabla_zH\Bigl(\frac{1}{2}(z^{(n+1)}
+ z^{(n)})\Bigr)
-\frac{h^3}{24}J^{-1}\nabla_z((\nabla_zH)^T
JH_{zz}J\nabla_zH)
\Bigl(\frac{1}{2}(z^{(n+1)}+ z^{(n)})\Bigr)\,.
\end{align}
In this case $\cal H$ can be introduced as 
\begin{align}
{\cal H} = H-\frac{h^2}{24}(\nabla_zH)^TJH_{zz}J\nabla_zH\,.
\end{align}
Then the fourth-order symplectic scheme can be rewritten as
\begin{align}
z^{(n+1)} = z^{(n)}+hJ^{-1}\nabla_z{\cal H}\Bigl(\frac{1}{2}
(z^{(n+1)}+ z^{(n)})\Bigr)\,.
\end{align}
Introducing an associated DEL form,
\begin{align}
{E_{D2}}^{(k)}=\frac12 \d({z^{(k+1)}+z^{(k)})}^T\Bigl\{ J \Delta_t z^{(k)}
-\nabla_z {\cal H}^{(k)}\Bigl(\frac12(z^{(k+1)}
+z^{(k)})\Bigr)\Bigr\}\,,
\end{align}
it is easy to see that $E_{D1}$ and $E_{D2}$ differ an exact form,
\begin{align}
{E_{D1}}^{(k)}-{E_{D2}}^{(k)}=\frac{h^2}{24}\d\alpha\,, 
\qquad  
\alpha=(\nabla_zH)^TJH_{zz}J\nabla_zH\,.
\end{align}

The second example is symplectic Runge--Kutta (RK) scheme. First, the 
stage one and order-two symplectic RK method is nothing but the mid-point 
scheme. Second, the stage two and order-four RK method is as follows:
\begin{align}\alw
& y^{(n+1)}=y^{(n)}+\frac {h}{2}(f(Y_1)+f(Y_2))\,,
\nonumber\\\noa2
& Y_1=y^{(n)}+h\Bigl[\frac {1}{4}f(Y_1)+\Bigl(\frac {1}{4}
+\frac {1}{2\sqrt 3}\Bigr)f(Y_2)\Bigr]\,,
\nonumber\\\noa2
& Y_2=y^{(n)}+h\Bigl[\Bigl(\frac {1}{4}-\frac {1}{2\sqrt 3}\Bigr)
f(Y_1)+\frac {1}{4}f(Y_2)\Bigr]\,.
\end{align} 
It can be expressed in terms of Hamiltonian $H$ as
\begin{align}
\alw 
& q^{(n+1)}=q^{(n)}+\frac {h}{2}[H_p(P_1,Q_1)+H_p(P_2, Q_2)]\,,
\nonumber\\\noa2
& p^{(n+1)}=p^{(n)}-\frac {h}{2}[H_q(P_1,Q_1)+H_q(P_2, Q_2)]\,,
\end{align} 
where
\begin{align}\alw
&Q_1=q^{(n)}+h\Bigl[\frac {1}{4}H_p(P_1,Q_1)+\Bigl(\frac {1}{4}+
\frac {1}{2\sqrt 3}\Bigr)H_p(P_2, Q_2)\Bigr]\,,
\nonumber \\\noa2
&P_1=p^{(n)}-h\Bigl[\frac {1}{4}H_q(P_1,Q_1)+\Bigl(\frac {1}{4}+
\frac {1}{2\sqrt 3}\Bigr)H_q(P_2, Q_2)\Bigr]\,,
\nonumber\\\noa2
&Q_2=q^{(n)}+h\Bigl[\Bigl(\frac {1}{4}-\frac {1}{2\sqrt 3}\Bigr)H_p(P_1,Q_1)
+\frac {1}{4}H_p(P_2, Q_2)\Bigr]\,,
\nonumber\\\noa2
&P_2=p^{(n)}-h\Bigl[\Bigl(\frac {1}{4}-\frac {1}{2\sqrt 3}\Bigr)H_q(P_1,Q_1)
+\frac {1}{4}H_q(P_2, Q_2)\Bigr]\,.
\end{align}
Introducing a pair of the DEL one-forms 
\begin{align}
&E_D^{(n)}(q):\;=\d p^{(n)}\Bigl\{q^{(n+1)}-q^{(n)}
-\frac {h}{2}(H_p(P_1,Q_1)+H_p(P_2, Q_2))\Bigr\}\,,
\nonumber\\\noa2
&E_D^{(n)}(p):\;=\d q^{(n)}\Bigl\{p^{(n+1)}-p^{(n)}
+\frac {h}{2}(H_q(P_1,Q_1)+H_q(P_2, Q_2))\Bigr\}\,, 
\end{align}
then the DEL conditions, i.e., their closed condition
\begin{align}
\d(E_D^{(n)}(q)+ E_D^{(n)}(p))=0\,,
\end{align}
gives rise to the symplectic preserving property
\begin{align}
\d p^{(n+1)}\wedge \d q^{(n+1)}=\d p^{(n)}\wedge \d q^{(n)}\,.
\end{align}
It can also be shown that $\omega^{(n+1)}=\d p^{(n+1)}\wedge \d
q^{(n+1)}$ may be expressed as  $\d q^{(n+1)}\wedge \d q^{(n)}$
with some coefficients.

\sect{The Trapezoidal Integrator}
\vspace*{-0.15cm}

It is well known that this scheme is good enough in comparison with other
well-known symplectic schemes. But for the long time, it is not clear why it
is so satisfactory. 

We will show that this scheme {\bfit is} symplectic, but the preserved 
symplectic structure is {\bfit not} simply $\omega=\d p^T \wedge \d q$. 
Of course, the preserved symplectic structure should be canonically 
transformed to the one in the simple form with different canonical 
coordinates and momenta in principle.

The scheme is given by
\begin{align}
&q^{(n+1)}-q^{(n)}=\frac {h}{2}[H_p^{(n+1)}(p^{(n+1)},q^{(n+1)})
+H_p^{(n)}(p^{(n)},q^{(n)})]\,,
\nonumber\\\noa2
&p^{(n+1)}-p^{(n)}=-\frac {h}{2}[H_q^{(n+1)}(p^{(n+1)},q^{(n+1)})
+H_q^{(n)}(p^{(n)},q^{(n)})]\,.
\end{align}

\subsect{For Separable Hamiltonian Systems}       

Let us now first consider the case of separable Hamiltonian systems.
For example $H=\frac {1}{2}p^2+V(q)$. In this case, the scheme reads
\begin{align}
q^{(n+1)}-q^{(n)}=\frac {h}{2}(p^{(n+1)}+p^{(n)})\,,\qquad
p^{(n+1)}-p^{(n)}=-\frac {h}{2}[V_q^{(n+1)}(q^{(n+1)})
+V_q^{(n)}(q^{(n)})]\,.
\end{align}
As what have been done before, let us introduce a pair of  the 
EL one-forms,
\begin{align}
&E_D^{(n)}(q):\;=\d p^{(n)}\Bigl\{q^{(n+1)}-q^{(n)}
-\frac {h}{2}(p^{(n+1)}+p^{(n)})\Bigr\}\,,
\nonumber\\\noa3
&E_D^{(n)}(p):\;=\d q^{(n)}\Bigl\{p^{(n+1)}-p^{(n)}
+\frac {h}{2}[V_q^{(n+1)}(q^{(n+1)})-V_q^{(n)}(q^{(n)}]\Bigr\}\,.     
\end{align}
Then by some straightforward but more or less tedious calculations, it follows
from the DEL condition, i.e. their closed property, that
$$\displaylines{\hfill
\frac {2}{h}(\d q^{(n)})^T\Bigl(1+\frac {h^2}{4}V_{qq}^{(n)}\Bigr)\wedge 
(\d q^{(n+1)}+\d p^{(n-1)})
=-\frac{h}{2}(\d q^{(n)})^T\Bigl(1+\frac {h^2}{4}V_{qq}^{(n)}\Bigr)
\wedge(V_{qq}^{(n+1)}\d q^{(n+1)}+V_{qq}^{(n-1)}\d q^{(n-1)}).
\hfill(62)\cr}
$$
We get
\setcounter{equation}{62}
\begin{align}
(\d q^{(n)})^T\Bigl(1+\frac {h^2}{4}V_{qq}^{(n)}\Bigr)
\wedge \Bigl(1+\frac {h^2}{4}V_{qq}^{(n+1)}\Bigr)\d q^{(n+1)}
=-(\d q^{(n)})^T\Bigl(1+\frac {h^2}{4}V_{qq}^{(n)}\Bigr)
\wedge \Bigl(1+\frac {h^2}{4}V_{qq}^{(n-1)}\Bigr)\d q^{(n-1)}\,.
\end{align}
This means that  the following symplectic structure  is preserved 
\begin{align}
(\d q^{(n+1)})^T\Bigl(1+\frac {h^2}{4}V_{qq}^{(n+1)}\Bigr)
\wedge \Bigl(1+\frac {h^2}{4}V_{qq}^{(n)}\Bigr)\d q^{(n)}\,.
\end{align}
That is
\begin{align}
(\d p^{(n+1)})^T \wedge \Bigl(1+\frac {h^2}{4}V_{qq}^{(n+1)}\Bigr)
\d q^{(n+1)}=(\d p^{(n)})^T\wedge \Bigl(1+\frac {h^2}{4}V_{qq}^{(n)}\Bigr)
\d q^{(n)}\,.
\end{align}
It is straightforward to show that this two-form is closed and
non-degenerate so that it is
the preserved symplectic structure for this scheme.

Using the following relation
\begin{align}
(\d p^{(n+1)})=\Bigl(1+\frac {h^2}{4}V_{qq}^{(n+1)}\Bigr)
\d q^{(n+1)}-\Bigl(1+\frac {h^2}{4}V_{qq}^{(n)}\Bigr)\d q^{(n)}\,,
\end{align}
one will get this two-form to be the same as Eq.~(64).

For the general separable Hamiltonian $H=T(p)+V(q)$ we can get the 
preserved symplectic structure for the scheme as follows:
\begin{align}
\omega^{(n+1)}=(\d p^{(n+1)})^T
\wedge \Bigl(1+\frac {h^2}{4}T_{pp}^{(n+1)}V_{qq}^{(n+1)}\Bigr)
\d q^{(n+1)}\,.
\end{align}
It is also closed and nondegenerate.

\subsect{The Trapezoidal Scheme for General Hamiltonian Systems}

For the general system with non-separable Hamiltonian, the trapezoidal scheme 
gives 
\begin{align}
q^{(n+1)}-q^{(n)}=\frac {h}{2}(H_p^{(n+1)}+H_p^{(n)}),\qquad
p^{(n+1)}-p^{(n)}=-\frac {h}{2}(H_q^{(n+1)}+H_q^{(n)})\,.
\end{align}
Similar to the separable Hamiltonian case, let us introduce a pair
of DEL one-forms,
\begin{align}\alw
&E_D^{(n)}(q):\;=\d p^{(n)}\Bigl\{q^{(n+1)}-q^{(n)}-
\frac {h}{2}(H_p^{(n+1)}+H_p^{(n)})\Bigr\}\,,
\nonumber \\\noa3
&E_D^{(n)}(p):\;=\d q^{(n)}\Bigl\{p^{(n+1)}-p^{(n)}+\frac {h}{2}(H_q^{(n+1)}
+H_q^{(n)})\Bigr\}\,.
\end{align}
Similarly, by some straightforward but tedious calculations, the DEL condition
for the pair of DEL one-forms gives rise to the following symplectic 
two-form and its preserving property,
\begin{align}
\omega_{\ssc D}^{(n+1)}=\omega_{\ssc D}^{(n)}\,,
\end{align}
where
\begin{align}
\omega_{\ssc D}^{(n)}=(\d p^{(n)})^T\Bigl(1+\frac {h^2}{4}H_{pp}^{(n)} 
H_{qq}^{(n)}-  
\frac {h^2}{4}H_{pq}^{(n)} H_{pq}^{(n)}-H_{pq}^{(n)} H_{pp}^{(n)}
\Bigr)\wedge \d q^{(n)}
-\frac {h^2}{4}(\d q^{(n)})^T H_{qq}^{(n)} H_{pq}^{(n)}\wedge \d q^{(n)}\,.
\end{align}
\begin{multicols}{2}
\noindent
If we introduce  new variables
\begin{align} 
& \tilde{p}^{(n)}= p^{(n)}-\frac {h}{2} H_{q}^{(n)}, \nonumber\\\noa2
& \tilde{q}^{(n)}= q^{(n)}+\frac {h}{2} H_{p}^{(n)}\,,
\end{align} 
it follows that
\begin{align}
\omega_{\ssc D}^{(n)}=\d\tilde{p}^{(n)}\wedge \d\tilde{q}^{(n)}\,.
\end{align}
This is another expression for the preserved symplectic structure in the 
trapezoidal scheme.

\sect{Some Remarks}
\begin{mylabellist}{iii)}
\item[i)] In order to show whether a scheme for a given Hamiltonian system is 
symplectic preserving, the first issue in our approach to be considered is to 
release the scheme from the solution space to the function space. 
Otherwise, it is difficult to make precise sense for the differential
calculation in the solution space.
 One of the roles played by the EL 
cohomological concepts is just to release the schemes from the solution 
space to the function space.
\item[ii)] The EL cohomology and its discrete counterpart introduced
in Ref.~[5] and used here are not trivial for the finite dimensional 
Hamiltonian systems. It has been shown that the symplectic
preserving property is closely linked to the cohomology. Namely,
it is equivalent to the closed condition of the EL one-forms. 
Of course, it is needed to further study the content and  meaning of
the EL cohomology.
\item[iii)] It should be mentioned that all issues studied in this note can 
be generalized to the case of difference discrete phase space for the 
separable Hamiltonian systems.\wen{3,4}
\end{mylabellist}

\nonumsect{Acknowledgment}

The authors would like to thank Prof.~M.Z. QIN for informing us after our 
relevant work has been finished that the symplectic preserving 
property for the trapezoidal scheme had been given by Wang.\wen{7} 
\end{multicols}

\refrule

\begin{multicols}{2}

\end{multicols}

\vfill
\end{document}